\documentstyle{article}

\input{psfig}

\begin{document}

\title{Measurement of electron-neutrino electron elastic scattering}

% repeat the \author\address pair as needed
\author{
L.B. Auerbach,$^8$ R.L. Burman,$^5$ D.O. Caldwell,$^3$ E.D. Church,$^1$ \\
J.B. Donahue,$^5$ A. Fazely,$^7$ G.T. Garvey,$^5$ R.M. Gunasingha,$^7$ 
R. Imlay,$^6$ \\
W.C. Louis,$^5$ R. Majkic,$^{8}$ A. Malik,$^6$ W. Metcalf,$^6$ 
G.B. Mills,$^5$ \\
V. Sandberg,$^5$ D. Smith,$^4$ 
I. Stancu,$^1$\footnote{now at University of Alabama, Tuscaloosa, AL 35487}
M. Sung,$^6$ 
R. Tayloe,$^5$\footnote{now at Indiana University, Bloomington, IN 47405} \\ 
G.J. VanDalen,$^1$ 
W. Vernon,$^2$ N. Wadia,$^6$ D.H. White,$^5$ S. Yellin$^3$\\
(LSND Collaboration) \\
$^1$ University of California, Riverside, CA 92521 \\
$^2$ University of California, San Diego, CA 92093 \\
$^3$ University of California, Santa Barbara, CA 93106 \\
$^4$ Embry Riddle Aeronautical University, Prescott, AZ 86301 \\
$^5$ Los Alamos National Laboratory, Los Alamos, NM 87545 \\
$^6$ Louisiana State University, Baton Rouge, LA 70803 \\
$^7$ Southern University, Baton Rouge, LA 70813 \\
$^8$ Temple University, Philadelphia, PA 19122 }

\date{\today}
\maketitle

\begin{abstract}
The cross section for the elastic scattering reaction 
$\nu_e+e^-\rightarrow\nu_e+e^-$ was measured by the Liquid Scintillator 
Neutrino Detector using a $\mu^+$ decay-at-rest $\nu_e$ beam 
at the Los Alamos Neutron Science Center.
The standard model of electroweak physics predicts a large destructive 
interference between the charge current and neutral current channels for this 
reaction.
The measured cross section, 
$\sigma_{\nu_ee^-}=[10.1\pm1.1(stat.)\pm1.0(syst.)]\times E_{\nu_e}$ 
(MeV) $\times10^{-45}$ cm$^2$, agrees well with standard model expectations.
The measured value of the interference parameter, 
$I=-1.01\pm0.13(stat.)\pm0.12(syst.)$, 
is in good agreement with the standard model expectation of $I^{SM}=-1.09$.
Limits are placed on neutrino flavor-changing neutral currents.
An upper limit on the muon-neutrino magnetic moment 
of $6.8\times10^{-10}\mu_{Bohr}$ is obtained using 
the $\nu_\mu$ and $\bar{\nu}_\mu$ fluxes from $\pi^+$ and $\mu^+$ decay.
\end{abstract}

\section{Introduction}

Neutrino-electron elastic scattering is a simple, purely leptonic weak process 
that can provide precise tests of the standard model (SM) 
of electroweak interactions. 
Measurements of the reactions $\nu_\mu+e^-\rightarrow\nu_\mu+e^-$ and 
$\bar{\nu}_\mu+e^-\rightarrow\bar{\nu}_\mu+e^-$ have been used to 
determine the vector and axial vector electron-$Z$ couplings, 
$g_V$ and $g_A$\cite{LAA90,JDOR89,PVIL94}. 
These reactions proceed solely via the neutral current (NC) channel. 
In contrast, the reaction $\nu_e+e^-\rightarrow\nu_e+e^-$ proceeds 
via both the charged current (CC) and neutral current channels. 
This reaction is of interest primarily 
because it is one of the few reactions for which the SM predicts 
a large destructive interference between these two channels. 
In this paper, we report a measurement of this reaction that is 
in good agreement with the SM and with the previous measurement\cite{RCA93}.

The differential cross sections for $\nu_\mu$ scattering on electrons 
can be written 
\begin{equation}
\frac{d\sigma^{\nu_\mu}}{dy}=\sigma_0[g_L^2+g_R^2(1-y)^2]  
\end{equation}
for $E_\nu\gg m_e$, 
where $\sigma_0=G_F^2s/4\pi,~s=2m_eE_\nu,~y=E_e/E_\nu,~g_L=g_V+g_A$ 
and $g_R=g_V-g_A$. 
In the SM, $g_V=-\frac{1}{2}+2\sin^2\theta_W$ and $g_A=-\frac{1}{2}$. 
The total cross section is then
\begin{equation}
\sigma^{\nu_\mu}=\sigma_0\large( g_L^2+\frac{g_R^2}{3}\large).
\end{equation}
For $\bar{\nu}_\mu$ scattering on electrons $g_L$ and $g_R$ are 
interchanged so that
\begin{equation}
\sigma^{\bar{\nu}_\mu}=\sigma_{0}\large( g_R^2+\frac{g_L^2}{3}\large).
\end{equation}

For scattering of $\nu_e$ on electrons the presence of 
the CC diagram results in a differential cross section of
\begin{equation}
\frac{d\sigma^{\nu_e}}{dy}=\sigma_0\large[(g_L+2)^2+g_R^2(1-y)^2\large]
\end{equation}
and a total cross section 
\begin{equation}
\sigma^{\nu_e}=\sigma_0[(g_L+2)^2+\frac{g_R^2}{3}].
\end{equation}
To make explicit the interference of the NC and CC, we rewrite this as
\begin{equation}
\sigma^{\nu_e}=\sigma^{CC}+\sigma^I+\sigma^{NC},
\end{equation}
where $\sigma^{CC}=4\sigma_0,~\sigma^{NC}=\sigma_0(g_L^2+\frac{g_R^2}{3})$ 
and $\sigma^I=2I\sigma_0$. 
In the SM $I=2g_L=-2+4\sin^2\theta_W$ and 
$\sigma^{NC}=\sigma_0(1-4\sin^2\theta_W+\frac{16}{3}\sin^4\theta_W)$. 
Assuming $\sin^2\theta_W=0.23$, we get $I=-1.1$.
Thus there is a substantial negative interference between the NC and CC terms.
Including radiation correction\cite{Marciano83} and 
retaining terms in the mass of electron in the cross section formula, 
we obtain $I^{SM}=-1.09$ and $\sigma^{NC}=0.37\sigma_0$.

The measured value of the interference term can be used to set limits on 
neutrino flavor-changing neutral currents (FCNC)\cite{LMSE75}. 
Such currents would not be detectable in neutrino experiments 
which measure a pure NC process since the flavor of 
the outgoing neutrino is not observed. 
The interference term, however, depends on the interference of 
the CC with the flavor conserving part of the NC.

Neutrino-electron elastic scattering can also be used to measure certain 
intrinsic electromagnetic properties of the neutrino. 
Neutrinos with magnetic moments will scatter electromagnetically on electrons. 
For electrons with recoil energy greater than $T$, 
the cross section for single-photon exchange dipole scattering is given by 
\begin{equation}
\label{eq:xsec_em}
\sigma^{EM}(E_\nu)=f^{2}{\pi}r^2_0[T/E_\nu-\ln(T/E_\nu)-1],
\end{equation}
where $r_0$ is the classical electron radius ($r_0=2.82\times10^{-13}$ cm) and 
$f$ is the ratio of the neutrino magnetic moment to the electron Bohr magneton.
The present experiment can set upper limits on both the $\nu_e$ 
magnetic moment and the $\nu_\mu$ magnetic moment 
because it observes scattering of $\nu_\mu$ and $\bar{\nu}_\mu$ 
as well as $\nu_e$ on electrons. 
However, only the upper limit obtained on the $\nu_\mu$ magnetic moment 
is competitive with previous laboratory limits\cite{LAA90,RCA93}.

We also place limits on anomalous contributions to the Lorentz and Dirac 
structure of the scattering amplitude, which would be manifested 
as changes in the electromagnetic form factors\cite{ROBE82} and, 
in particular, as an effective neutrino charge radius\cite{GDEG89}. 
The weak NC coupling, $g_V$, would shift to $g_V+2\delta$ 
where $\delta=(\sqrt{2}\pi\alpha/3G_F)\langle r^2\rangle
=(2.39\times10^{30}$ cm$^{-2})\langle r^2\rangle$ and $\langle r^2\rangle$ 
is the effective mean squared charge radius of the neutrino. 
As defined, $\langle r^2\rangle$ is gauge dependent and 
not necessarily positive.
It provides a procedure, however, for paramaterizing certain not-standard 
contributions to neutrino scattering.
Our measurement of $g_V$ in elastic scattering provides a limit on internal 
electromagnetic structure at the level of $10^{-16}$ cm.

The interference of the NC ($Z$ exchange) and CC ($W$ exchange) terms 
can be studied in several other reactions, including 
$\nu_\mu N\rightarrow\nu_\mu\mu^+\mu^-N,~\bar{\nu}_ee^-\rightarrow\bar{\nu}_ee^-$ 
and $e^+e^-\rightarrow\gamma\nu\bar{\nu}$. 
Neutrino trident production, $\nu_\mu N\rightarrow\nu_\mu\mu^+\mu^-N$, 
was first clearly observed by the CHARM II experiment, 
but this experiment was not sensitive enough to demonstrate destructive 
interference\cite{DGEI90}. 
CCFR saw destructive interference and was able to rule out 
a pure $W$ exchange interaction\cite{SRMI91}. 
A more recent measurement by the NuTeV experiment, however, 
could not distinguish between $W$ exchange alone and the SM\cite{TADA00}. 
Further, they note that earlier analyses of trident production 
did not consider diffractive sources. 
The cross section for $\bar{\nu}_e+e^-\rightarrow\bar{\nu}_e+e^-$ 
was measured in a reactor experiment and found to be in agreement 
with the SM but also consistent, within errors, 
with a CC interaction\cite{FREI77}.

The reaction $e^+e^-\rightarrow\gamma\nu\bar{\nu}$ has been extensively 
studied at LEP \cite{LEP2}. 
Near the $Z^0$, the NC term dominates and the interference term is small. 
At higher energies the interference term can contribute as much as $25\%$, 
but is also sensitive to the event selection criteria used. 
The LEP2 measurements agree well with the SM and have been used 
to set limits on various possible new physics processes, 
but we are not aware of any explicit measurements of the interference term.

\section{THE NEUTRINO SOURCE}
\label{sec:source}

\begin{figure}
\centerline{\psfig{figure=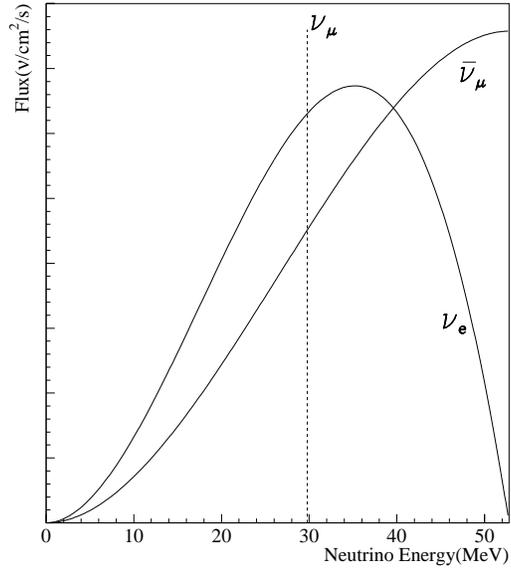,height=3in}}
\caption{Flux shape of neutrinos from pion and muon decay at rest.}
\label{fig:flux}
\end{figure}

The data reported here were obtained between 1994 and 1998 
by the Liquid Scintillator Neutrino Detector (LSND) at the Los Alamos 
Neutron Science Center (LANSCE)
using neutrinos produced at the A6 proton beam stop.
The neutrino source is described in detail elsewhere\cite{Athan97}.
In 1994 and 1995 the beam stop consisted of a 30 cm water target 
surrounded by steel shielding and followed by a copper beam dump.
The high-intensity 798 MeV proton beam from the linear accelerator 
generated a large pion flux from the water target.
The flux of $\nu_e$ used for the measurements reported here arise 
from the decay at rest (DAR) of stopped $\pi^+$ and $\mu^+$.
This decay chain yields almost equal intensities of $\nu_e,~\bar{\nu}_\mu$ and 
$\nu_\mu$ with the well-determined energy spectra shown 
in Figure \ref{fig:flux}.

After the 1995 run the beam stop was substantially modified for accelerator 
production of tritium (APT) tests.
The most significant change for the analysis presented in this paper was the 
replacement of the water target by tungsten and other materials with high
atomic number.
This resulted in reduced $\pi^+$ production and a lower DAR neutrino flux, 
largely due to the change in the neutron to proton ratio in the target.

The corresponding decay chain for $\pi^-$ and $\mu^-$ is highly suppressed 
due to three factors. 
First, production of $\pi^-$ is smaller than for $\pi^+$. 
Second, $\pi^-$ which stop are absorbed by nuclear interactions. 
Finally, most $\mu^-$ which stop are absorbed before they can decay. 
These stopped $\mu^-$ arise from $\pi^-$ which decay in flight (DIF).

The LANSCE beam dump has been used as the neutrino source for previous 
experiments\cite{Willis80,Krak92,Free93}.
A calibration experiment\cite{All89} measured the rate of stopped $\mu^+$ 
from a low-intensity proton beam incident on an instrumented beam stop.
The rate of stopped $\mu^+$ per incident proton was measured as a function of 
several variables and 
used to fine-tune a beam dump simulation program\cite{Bur90}.
The simulation program can then be used to calculate the flux 
for any particular beam dump configuration.
The calibration experiment determined the DAR flux to $\pm7\%$ for the proton 
energies and beam stop configurations used at LANSCE.
This uncertainty provides the largest source of systematic error for 
the cross sections presented here.
The LANSCE proton beam typically had a current of 800 $\mu$A at the beam stop. 
For 1994 and 1995 the energy was approximately 770 MeV at the beam stop 
due to energy loss in upstream targets.
The integrated beam current was 5904 C in 1994 and 7081 C in 1995.
The calculated ratio of stopped $\mu^+$ per proton was 0.090 and 0.084 
for 1994 and 1995, respectively, with the lower ratio for 1995 arising 
because the water target was out for $32\%$ of the 1995 data.
Upstream targets contributed 1.4$\%$ to the DAR flux in 1994 and 1995.
The DAR $\nu_e$ flux averaged over the LSND detector was then 
$3.08\times10^{13}$ cm$^{-2}$ for 1994 and $3.45\times10^{13}$ cm$^{-2}$ 
for 1995.

The 1996-1998 data was obtained with the APT beam stop.
There were no upstream targets for almost all of the data taking.
The integrated beam current was 3789 C in 1996, 7181 C in 1997 and 3155 C 
in 1998.
The calculated ratio of stopped $\mu^+$ per incident proton was 
0.069, 0.068 and 0.067 respectively in 1996, 1997 and 1998.
The DAR $\nu_e$ flux average over the LSND detector was 
$1.32\times10^{13}$ cm$^{-2}$ for 1996, $2.73\times10^{13}$ cm$^{-2}$ 
for 1997 and $1.18\times10^{13}$ cm$^{-2}$ for 1998. 
For the full data sample used in this paper the $\nu_e$ flux 
is $11.76\times10^{13}$ cm$^{-2}$.

\section{The LSND Detector}
\label{sec:lsnd}

\begin{figure}
\centerline{\psfig{figure=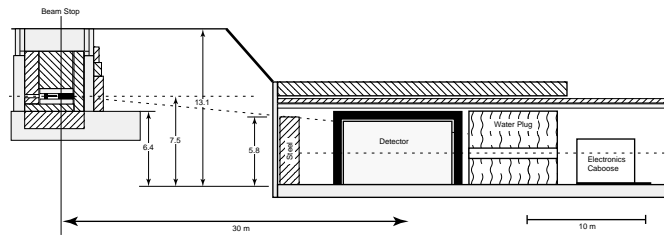,height=1.2in}}
\caption{Detector enclosure and target area configuration, elevation view.}
\label{fig:detector}
\end{figure}

The detector is located 29.8 m downstream of the proton beam stop 
at an angle of $12^\circ$ to the proton beam.  
Figure \ref{fig:detector} shows a side-view of the setup.  
Approximately 2000 g/cm$^2$ of shielding above the detector attenuates 
the hadronic component of cosmic rays to a negligible level.  
The detector is also well shielded from the beam stop 
so that beam-associated neutrons are attenuated to a negligible level.  
Enclosing the detector, except on the bottom, 
is a highly efficient liquid scintillator veto shield 
which is essential to reduce contributions 
from the cosmic ray muon background to a low level.    
Reference \cite{Athan97} provides a detailed description of the detector,
veto, and data acquisition system which we briefly review here.  

The detector is an approximately cylindrical tank containing 167 tons of 
liquid scintillator and viewed by 1220 uniformly spaced $8''$ Hamamatsu 
photomultiplier tubes (PMT) covering $25\%$ of the surface 
inside the tank wall. 
When the deposited energy in the tank exceeds a threshold of 
approximately 4 MeV electron-equivalent energy and 
there are fewer than 4 PMT hits in the veto shield, 
the digitized time and pulse height of each of these PMTs 
(and of each of the 292 veto shield PMTs) are recorded.  
A veto, imposed for 15.2 $\mu$s following the firing of $>5$ veto PMTs, 
substantially reduces ($10^{-3}$) 
the large number of background events arising from the decay 
of cosmic ray muons that stop in the detector.  
Activity in the detector or veto shield during the 51.2 $\mu$s 
preceding a primary trigger is also recorded 
provided there are $>17$ detector PMT hits or $>5$ veto PMT hits. 
This activity information is used in the analysis to reject events 
arising from muon decay. 
Data after the primary event are recorded for 1 ms with a threshold 
of 21 PMTs (approximately 0.7 MeV electron-equivalent energy). 
This low threshold is used for the detection of 2.2 MeV $\gamma$ 
from neutron capture on free protons.  
In the present analysis this information is used to help identify events 
induced by cosmic ray neutrons.  
The detector operates without reference to the beam spill, 
but the state of the beam is recorded with the event.  
Approximately $94\%$ of the data is taken between beam spills.  
This allows an accurate measurement and subtraction of cosmic ray background 
surviving the event selection criteria.  

Most triggers due to electrons from 
stopped muon decays (Michel electrons) are identified 
by a preceding activity produced by the decay muon.  
Ocassionally, however, the muon will not satisfy 
the activity threshold of $>17$ detector PMT hits or $>5$ veto PMT hits.  
For several LSND analyses, including the present one, 
it is desirable to further reduce the number of unidentified Michel electrons. 
Therefore, for data recorded after 1994 all PMT information was recorded 
for a period of 6 $\mu$s (2.7 muon lifetimes) preceding certain primary events.
This ``lookback" information was recorded for primary events 
with $>300$ PMT hits and no activity within the past 35 $\mu$s (20 $\mu$s) 
for 1995 data (1996-1998 data). 
For the present analysis this ``lookback" information is used 
to further reduce the cosmic ray muon background.  

The detector scintillator consists of mineral oil ($CH_2$) in 
which is dissolved a small concentration (0.031 g/l) of b-PBD\cite{Ree93}. 
This mixture allows the separation of \v{C}erenkov light and 
scintillation light and produces about 33 photoelectrons per MeV of 
electron energy deposited in the oil.  
The combination of the two sources of light provides direction information 
for relativistic particles and makes particle identification (PID) possible.
Note that the oil consists almost entirely of carbon and hydrogen.  
Isotopically the carbon is $1.1\%~^{13}C$ and $98.9\%~^{12}C$.

The veto shield encloses the detector on all sides except the bottom.  
Additional counters were placed below the veto shield 
before the 1994 run to reduce cosmic ray background entering 
through the bottom support structure.  
These counters around the bottom support structure are referred to 
as bottom counters.  
More bottom counters were added after the 1995 run.
The main veto shield\cite{Nap89} consists of a 15-cm layer of liquid 
scintillator in an external tank and 15 cm of lead shot in an internal tank.  
This combination of active and passive shielding tags cosmic ray muons 
that stop in the lead shot.  
A veto inefficiency $<10^{-5}$ is achieved with this detector 
for incident charged particles.  

\section{Analysis Techniques}
\label{sec:analysis}

Each event is reconstructed using the hit time and pulse height of all hit 
PMTs in the detector\cite{Athan97}.
The present analysis relies on the reconstructed energy, position, direction, 
and two PID parameters, $\chi'_{tot}$ and $\alpha$, 
as described later in this section.
The particle direction is determined from the \v{C}erenkov cone.
The parameters $\chi'_{tot}$ and $\alpha$ are used to distinguish electron 
events from events arising from interactions of cosmic ray neutrons 
in the detector.
Fortunately, it is possible to directly measure the response of the detector 
to electrons and neutrons in the energy range of interest for this analysis 
by using copious control data samples.
We also use a Monte Carlo simulation, LSNDMC\cite{McI95}, to simulate events 
in the detector using GEANT.

The response of the detector to electrons was determined from a large, 
essentially pure sample of electrons (and positrons) from the decay of 
stopped cosmic ray $\mu^\pm$ in the detector.
The known energy spectra for electrons from muon decay was used 
to determine the absolute energy calibration, 
including its small variation over the volume of the detector.
The energy resolution was determined from the shape of the electron energy 
spectrum and was found to be $6.6\%$ at the 52.8 MeV end-point.
The position and direction resolution obtained from the LSNDMC simulation 
are 27 cm and $17^\circ$, respectively, for electrons from $\nu e$ 
elastic scattering in the energy region above 18 MeV.
The accuracy of the direction measurement is discussed more in 
Section \ref{sec:elastic} since the measurement of the angular distribution 
of electrons is crucial for the analysis presented in this paper. 
Electrons from $\nu e$ elastic scattering are sharply peaked 
along the incident neutrino direction, 
while electrons from other neutrino processes have 
a broad angular distribution that peaks in the backward direction.

There are no tracking devices in the LSND detector. 
Thus, event positions must be determined solely from the PMT information.
The reconstruction process determines an event position by minimizing 
a function $\chi_r$ which is based on the time of each PMT hit corrected for 
the travel time of light from the assumed event position 
to the PMT\cite{Athan97}.
The procedure used in several previous analyses systematically shifted 
event positions away from the center of the detector and thus 
effectively reduced the fiducial volume\cite{At96}.
The reconstruction procedure has been analyzed in detail and 
an improved reconstruction procedure was developed which reduces 
this systematic shift and provides substantially better position resolution.
This procedure also provides results which agree well with positions 
obtained from the event likelihood procedure described in Ref. \cite{At98}.
In the analysis presented in this paper, a fiducial cut is imposed 
by requiring $D>35$ cm, where $D$ is the distance between the reconstructed 
event position and the surface tangent to the faces of the PMTs. 
Events near the bottom of the detector ($y<-120$ cm) are also removed, 
as discussed in Section \ref{sec:electron}.

The particle identification procedure is designed to separate particles 
with velocities well above \v{C}erenkov threshold from particles 
below \v{C}erenkov threshold.
The procedure makes use of the four parameters defined in Ref. \cite{Athan97}.
Briefly, $\chi_r$ and $\chi_a$ are the quantities minimized for 
the determination of the event position and direction, $\chi_t$ is 
the fraction of PMT hits that occur more than 12 ns after the fitted 
event time and $\chi_{tot}$ is proportional to the product of $\chi_r$, 
$\chi_a$ and $\chi_t$.

Several previous LSND analyses \cite{At96,At97,Ath97} have used $\chi_{tot}$ 
for particle identification.
The distribution of $\chi_{tot}$ for electrons, however, has a small variation 
with electron energy and with the position of the event.
Therefore, in this paper, we used a modified variable, $\chi'_{tot}$, with 
a mean of zero and sigma of one, independent of the electron energy and 
positions. 
We also used the variable, $\alpha$, which is based on the event 
likelihood procedures discussed in Ref. \cite{At98}.
It is similar to the parameter $\rho$ discussed there, 
which is based on the ratio of \v{C}erenkov to scintillator light.
The parameter $\alpha$ varies from 0 to 1 and 
peaks at one for electrons and at 0.3 for neutrons.
The combination $\chi_\alpha=\chi_{tot}^\prime+10(1-\alpha)$ provides better 
separation of electrons and neutrons than $\chi_{tot}^\prime$ by itself.

\begin{figure}
\centerline{\psfig{figure=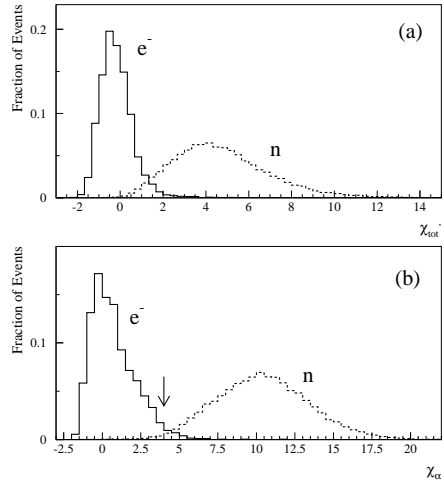,height=2.8in}}
\caption{Particle identification parameters (a) $\chi_{tot}\prime$ and 
(b) $\chi_\alpha$ for electrons and neutrons.
In the present analysis we require $\chi_\alpha<4.0$ as indicated by the 
arrow in (b).}
\label{fig:eid}
\end{figure}

Figure \ref{fig:eid}(a) shows the $\chi'_{tot}$ distribution for electrons 
from stopping $\mu$ decay and for cosmic ray neutrons with electron equivalent 
energies in the $18<E_e<50$ MeV range. 
Neutrons, after thermalizing, undergo a capture reaction, 
$n+p\rightarrow d+\gamma$. 
The 2.2 MeV $\gamma$'s are used to select a clean sample of neutron events. 
For a neutron $E_e$ is the equivalent electron energy corresponding to 
the observed total charge.
Figure \ref{fig:eid}(b) shows the $\chi_\alpha$ distribution for 
the same events.
In the present analysis we eliminate most cosmic ray neutron background by 
requiring $\chi_\alpha<4.0$. 
We note that a modest particle identification requirement was imposed 
in the initial data processing that created the samples analyzed here. 
The effect of this requirement is also included in the analysis.

Beam-off data taken between beam spills play a crucial role in the analysis 
of this experiment.
Most event selection criteria are designed to reduce the cosmic ray background 
while retaining high acceptance for the neutrino process of interest.
Cosmic ray background which remains after all selection criteria 
have been applied is well measured with the beam-off data and 
subtracted using the duty ratio, the ratio of beam-on time to beam-off time.
The subtraction was performed separately for each year's data 
using the measured duty ratio for that year.
The ratio averaged over the data full sample was 0.0632.
Beam-on and beam-off data have been compared to determine if there are 
any differences other than those arising from neutrino interactions.
Any differences are small and the $1.1\%$ uncertainty in the duty ratio 
introduces a negligible effect in the present analysis.

\begin{figure}
\centerline{\psfig{figure=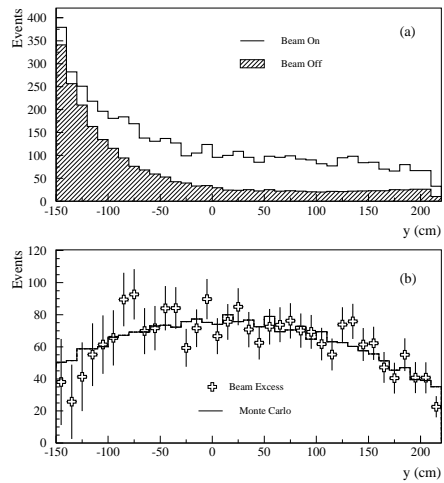,height=2.8in}}
\caption{The $y$ distribution of inclusive electrons (a) for beam-on events 
and for beam-off events corrected by the duty ratio (cross hatched), and 
(b) beam-excess events compared with Monte Carlo expectations (solid line).}
\label{fig:ply}
\end{figure}

The beam-off subtraction procedure is illustrated in Figure \ref{fig:ply} 
for the $y$ distribution of the sample of inclusive electron events discussed 
in the next section.
Figure \ref{fig:ply}(a) shows the $y$ distribution for beam-on events and 
for beam-off events corrected by the duty ratio.
The beam-off background is largest at low $y$ due to the absence of a veto 
below the detector.
Figure \ref{fig:ply}(b) compares the $y$ distribution of the beam-excess 
events with that expected from neutrino processes.
The agreement is excellent.

\section{Inclusive Electron Sample}
\label{sec:electron}

Beam-associated electrons below 52 MeV in LSND arise from 
four major neutrino processes: 
$^{12}C(\nu_e,e^-)^{12}N_{g.s.},~^{12}C(\nu_e,e^-)^{12}N^*,
~^{13}C(\nu_e,e^-)^{13}X$ and $\nu e$ elastic scattering.
We distinguish transitions to the ground states ($^{12}N_{g.s.}$) and excited 
states ($^{12}N^*$) of nitrogen because the ground state has a clear 
signature from its $\beta$-decay. 
In this section we describe the selection criteria used to obtain 
a clean sample of  inclusive electron events arising from neutrino 
interactions in the detector.
The next section describes how the angular distribution of these electrons 
is used to obtain a $\nu e$ elastic sample and, in addition, 
to determine the background due to other neutrino processes 
remaining in that sample.

The selection criteria and corresponding efficiencies for electrons
from $\nu e$ elastic scattering are shown in Table \ref{ta:elecsel}.
A lower limit on the electron energy of 18.0 MeV eliminates 
the large cosmic ray background from $^{12}B~\beta$-decay 
as well as most 15.1 MeV gamma rays from the NC excitation of carbon.
The $^{12}B$ nuclei arise from the absorption of stopped $\mu^-$ on $^{12}C$ 
nuclei in the detector.
The requirement $y>-120$ cm removes a small region 
at the bottom of the detector for which the cosmic ray background is 
relatively high, as shown in Figure \ref{fig:ply}(a).
The reconstructed electron position is also required to be 
a distance $D>35$ cm from the surface tangent to the faces of the PMTs.
There are $2.72\times10^{31}$ electrons within this fiducial volume.
We show in the next section that we are able to make a good measurement 
of the direction for electrons within this fiducial volume.
The fiducial volume efficiency, defined to be the ratio of 
the number of events reconstructed within the fiducial volume 
to the actual number within this volume, was determined 
to be $0.918\pm0.055$. This efficiency is less than one because there 
is a systematic shift of reconstructed event positions away from the 
center of the detector as discussed in Section \ref{sec:analysis}.

Several selection criteria are designed to further reject cosmic ray 
induced events.
Events with more than three veto PMT hits or any bottom counter coincidence 
during the 500 ns event window are eliminated.
The past activity cut is designed to reject most electron events 
arising from cosmic ray muons which stop in the detector and decay.
This background has a time dependence  given by the 2.2 $\mu$s muon lifetime.
The past activity selection criteria reject all events with activity within 
the past 20 $\mu$s with $>5$ veto PMT hits or $>17$ detector PMT hits.
We also reject any event with a past activity within 51 $\mu$s 
with $>5$ veto PMT hits and $>500$ detector PMT hits. 
A small ($0.5\%$) loss of efficiency arises from a cut 
(made during initial data processing) on past activities 
that are spatially correlated with the primary event, 
within 30 $\mu$s of the primary event and have $\geq4$ veto PMT hits.

Muons which are misidentified as electrons are removed by requiring 
that there be no future activity consistent with a Michel electron.
Any event with a future activity with fewer than 4 veto PMT hits and 
more than 50 detector PMT hits within 8.8 $\mu$s is rejected.

\begin{figure}
\centerline{\psfig{figure=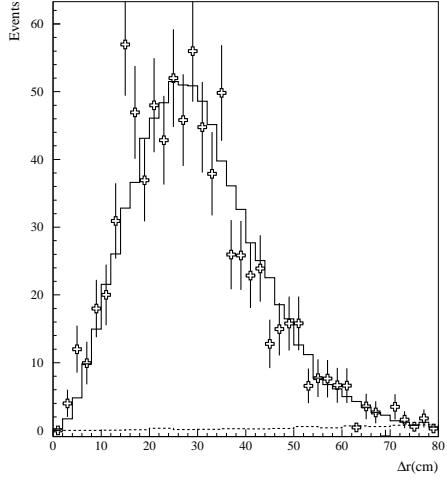,height=2.8in}}
\caption{Distribution of the distance between reconstructed positions of $e^-$ 
and $e^+$ for beam-excess events in the $^{12}C(\nu_{e},e^-)^{12}N_{g.s.}$ 
sample compared with Monte Carlo expectations (solid line). 
The calculated accidental contribution is shown by the dashed line.}
\label{fig:beta4}
\end{figure}

Electrons from the reaction $^{12}C(\nu_e,e^-)^{12}N_{g.s.}$ can 
be identified by the positron from the $\beta$-decay of the $^{12}N_{g.s.}$. 
Figure \ref{fig:beta4} shows the distance between the reconstructed 
electron and positron positions.
Table \ref{ta:betaeff} shows the $\beta$ selection criteria and 
corresponding efficiencies.
Reference \cite{At97} discusses in detail the measurement of the reaction 
$^{12}C(\nu_e,e^-)^{12}N_{g.s.}$.
For the $\nu_ee$ analysis we reject events with an identified $\beta$.

Cosmic ray muons which fire $<$6 veto PMTs ($10^{-3}$ probability) and 
stop in the iron walls of the detector will not register as past activities.
Some of the decay electrons will radiate photons 
which will enter the detector and be reconstructed 
as electrons within the fiducial volume.
In previous analyses we simply relied on the beam-off subtraction procedure 
to remove this background.
Here we use the ``lookback'' information described in Section \ref{sec:lsnd} 
to reject events from this source.
This results in slightly smaller statistical errors 
in the final beam-excess sample.

\begin{figure}
\centerline{\psfig{figure=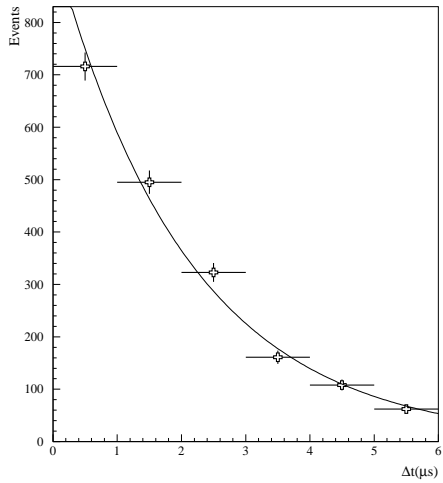,height=2.8in}}
\caption{The distribution of time between the primary and the veto signal for 
beam-off events rejected using the ``lookback'' information compared 
with a curve corresponding to the muon lifetime.}
\label{fig:beta5}
\end{figure}

For primary events with $>300$ PMT hits and no activity 
within the past 35 $\mu$s (20 $\mu$s) for 1995 data (1996-1998 data),
we recorded all PMT information for the 6 $\mu$s interval preceding the event.
Muons with $<6$ veto PMT hits will appear in this ``lookback'' interval 
as a cluster of veto PMT hits spatially correlated with the primary event.
The distribution of time between the veto signals and the primary event 
should be consistent with the muon lifetime, 
and the distributions of veto PMT hits and veto pulse height 
should be consistent with that measured for muons producing $<6$ veto PMT hits.
We developed a likelihood procedure based on these distributions 
which allowed us to reduce the beam-off background by $15\%$ 
with only a $0.6\%$ loss of efficiency for neutrino events\cite{Wad98}.
Figure \ref{fig:beta5} shows the time between the veto signal and 
the primary for rejected events.

The acceptances for the past activity, the future activity, 
the ``lookback'' and the in-time veto cuts are obtained by applying these cuts 
to a large sample of random events triggered with the laser used 
for detector calibration.
These laser events are spread uniformly through the run and 
thus average over the small variation in run conditions.
The acceptance for the 15.1 $\mu$s trigger veto is included 
in the past activity efficiency.

\begin{figure}
\centerline{\psfig{figure=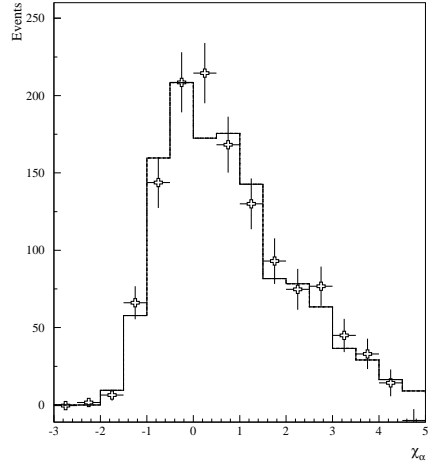,height=2.7in}}
\caption{The $\chi_\alpha$ distribution of the beam-excess inclusive 
electron sample. 
The histogram shows the $\chi_\alpha$ distribution of Michel electrons 
weighted and normalized to the same area.}
\label{fig:chialpha}
\end{figure}

A sample of Michel electrons 
was analyzed to obtain the acceptance of electrons for the PID cut.
Figure \ref{fig:chialpha} compares the $\chi_\alpha$ distribution of 
the inclusive electron sample with a weighted Michel electron sample.
The agreement is excellent.
To eliminate any energy dependence, the Michel electrons are given weights 
as a function of energy so that the weighted spectrum agrees with 
the energy spectrum of $\nu_e e$ elastic scattering.
The acceptance, however, is very insensitive to the assumed energy spectrum.
The beam-excess distribution shown in Figure \ref{fig:chialpha} is obtained 
by subtracting the beam-off distribution from the beam-on distribution 
as discussed in Section \ref{sec:lsnd}.

\section{Elastic Electron Sample}
\label{sec:elastic}

\begin{figure}
\centerline{\psfig{figure=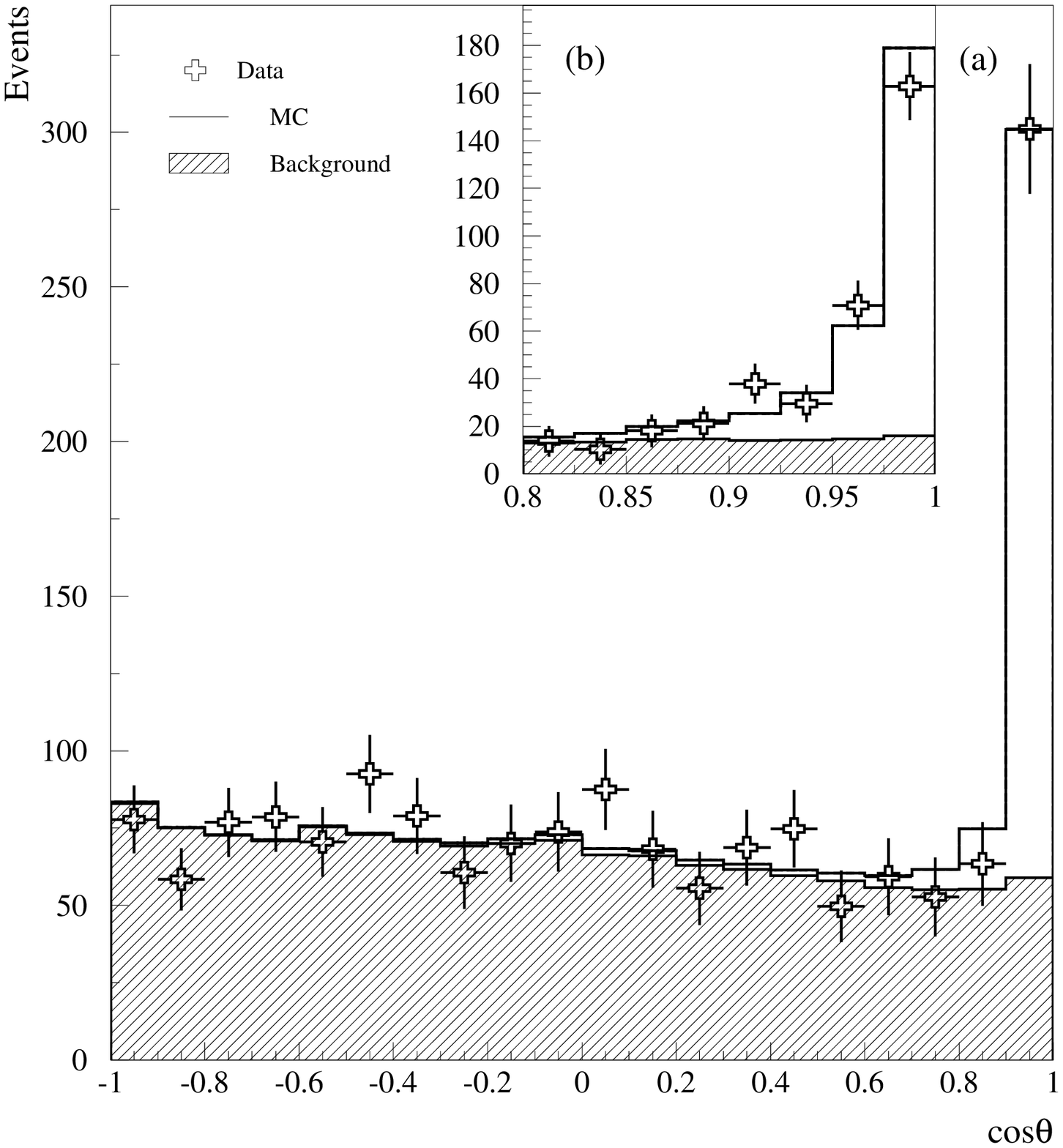,height=2.7in}}
\caption{The observed $\cos\theta$ distribution of the beam-excess inclusive 
electron sample (a) for all angles and (b) for $\cos\theta>0.8$. 
The solid line shows a fit to the data by the method explained in the text.
The hatched histogram shows the estimated background level 
from neutrino reactions other than $\nu e$ elastic scattering.}
\label{fig:plcos}
\end{figure}

\begin{figure}
\centerline{\psfig{figure=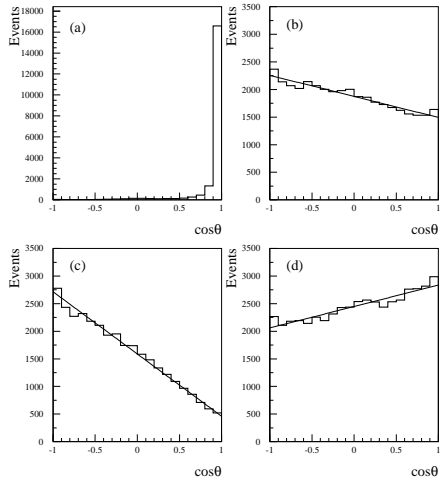,height=2.8in}}
\caption{The expected distributions of $\cos\theta$ for 
(a) $\nu e^-$ elastic scattering, (b) $^{12}C(\nu_e,e^-)^{12}N_{g.s.}$, 
(c) $^{12}C(\nu_e,e^-)^{12}N^*$, and (d) $^{13}C(\nu_e,e^-)^{13}X$.
Straight line fits are shown in (b), (c) and (d).}
\label{fig:plcos_mc}
\end{figure}

Figure \ref{fig:plcos} shows the observed distribution in $\cos\theta$ for 
the beam-excess inclusive electron sample, 
where $\theta$ is the angle between the reconstructed electron direction 
and the incident neutrino direction.
The large forward peak arises from $\nu e$ elastic scattering.
Figure \ref{fig:plcos_mc} shows the expected distributions in $\cos\theta$ 
for the primary sources of electrons in the sample: 
$\nu e$ elastic scattering, $^{12}C(\nu_e,e^-)^{12}N_{g.s.}$, 
$^{12}C(\nu_e,e^-)^{12}N^*$, and $^{13}C(\nu_e,e^-)^{13}X$.
These distributions have been obtained with the LSNDMC simulation package 
\cite{McI95} and thus include the angular smearing 
due to experimental effects. 
Before smearing, the expected distribution for all processes other than 
$\nu e$ elastic scattering varies gradually with $\cos\theta$, 
and a function of the form $A+B\cos\theta$ provides a good fit 
to the distribution. 
As seen in Figure \ref{fig:plcos_mc}, there is only a small deviation from 
linearity after experimental smearing.
Most $\nu e$ elastic events satisfy the selection criteria $\cos\theta>0.9$. 
The background from neutrino carbon scattering under the elastic peak 
($\cos\theta>0.9$) is obtained by fitting the observed distribution in 
Figure \ref{fig:plcos} to the sum of a term with the shape expected  
for $\nu e$ elastic scattering shown in Figure \ref{fig:plcos_mc} and 
a background term which differs slightly from the form $A+B\cos\theta$ 
due to the experimental smearing. % and 
From the fit we calculate a background of $59\pm5$ events 
with $\cos\theta>0.9$.
Measurements of $\nu_e C$ scattering, including angular distributions 
for $^{12}C(\nu_e,e^-)^{12}N_{g.s.}$ and $^{12}C(\nu_e,e^-)^{12}N^*$, 
will be reported in a separate paper \cite{newC}. 

\begin{figure}
\centerline{\psfig{figure=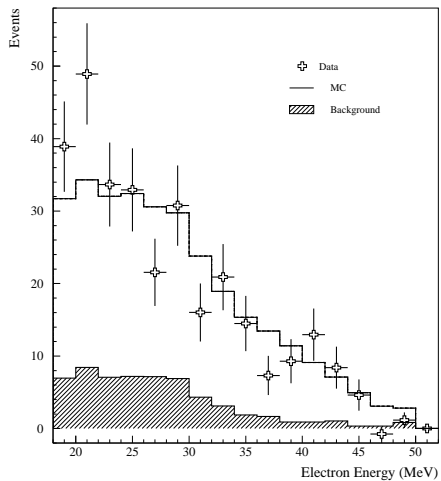,height=2.8in}}
\caption{The observed and expected (solid line) distributions of beam-excess 
events with $\cos\theta>0.9$. 
The expected distribution includes the estimated contribution from $\nu C$ 
(cross hatched) as well as $\nu e$ elastic scattering.}
\label{fig:e50}
\end{figure}

\begin{figure}
\centerline{\psfig{figure=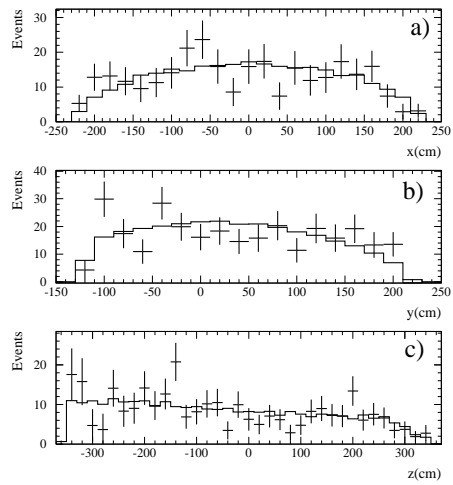,height=2.8in}}
\caption{The spatial distribution of the electron for beam-excess events with 
$\cos\theta>0.9$ compared with expectation (solid line) 
from $\nu e$ elastic scattering and $\nu C$ scattering.}
\label{fig:xyz}
\end{figure}

Figure \ref{fig:e50} shows the observed and expected electron energy 
distributions for beam-excess events with $\cos\theta>0.9$.
Figure \ref{fig:xyz} shows the observed and expected spatial distributions 
of the same events. 
Both figures show good agreement with expectations. 
Table \ref{ta:events} provides a breakdown of the number of events 
with $\cos\theta>0.9$, the calculated backgrounds, the acceptance, 
the neutrino flux and the resulting flux-averaged cross section 
for $\nu_ee$ elastic scattering. 
The dominant sources of systematic error in the cross section 
are the neutrino flux ($7\%$) discussed in Section \ref{sec:source}, 
the effective fiducial volume ($6\%$) discussed in Section \ref{sec:analysis}, 
the neutrino background with $\cos{\theta}>0.9~(3\%$), 
particle identification ($2\%$), the energy scale ($2\%$) and 
the direction determination ($2\%$).

All three types of DAR neutrinos ($\nu_e,~\nu_\mu$, and $\bar{\nu}_\mu$) 
elastically scatter off electrons in the detector, 
but the rate is dominated by $\nu_ee^-$ scattering\cite{K79}.
The contribution due to DIF $\nu_\mu$ and $\bar{\nu}_\mu$ scattering 
on electrons is small, approximately 6 events from $\nu_\mu$ scattering 
and $<1$ event from $\bar{\nu}_\mu$ scattering.
For the $\nu_e$ electron elastic scattering analysis, 
events from $\nu_\mu$ and $\bar{\nu}_\mu$ scattering are background and 
thus we subtract their contributions, shown in Table \ref{ta:events}, 
from the observed elastic scattering signal. 
Other experiments have measured cross sections for both $\nu_\mu$ 
and $\bar{\nu}_\mu$ scattering that are in good agreement 
with expectations\cite{Vil89}. 
The numbers in Table \ref{ta:events} for $\nu_\mu$ and $\bar{\nu}_\mu$ are 
obtained from the theoretical cross sections rather than the measured ones, 
although the analysis is insensitive to this choice. 
We also note that the sum of the contributions from $\nu_\mu$ and 
$\bar{\nu}_\mu$ scattering depends only weakly on the value 
of $\sin^2\theta_W$.

\section{Physics Results}
\label{sec:result}

The measured cross section, with its explicit linear energy dependence, is
\begin{equation}
\sigma_{\nu_ee^-}=[10.1\pm1.1(stat.)\pm1.0(syst.)]\times 
E_{\nu_e}({\rm MeV})\times10^{-45} {\rm cm}^2.
\end{equation}
This agrees well with the value measured by E225 at Los Alamos \cite{RCA93},
\begin{equation}
\sigma_{\nu_ee^-}=[10.0\pm1.5(stat.)\pm0.9(syst.)]\times 
E_{\nu_e}({\rm MeV})\times10^{-45} {\rm cm}^2
\end{equation} 
and with SM expectations.
The effective electron-$Z$ couplings, including ratiative corrections, are
$g_V=-0.0397$ and $g_A=-0.5064$ in the SM\cite{PDG98}. 
We then obtain, retaining terms in $m_e$ 
in the cross section formula\cite{Marciano83,tHooft71,Passera00}, 
\begin{equation}
\sigma_{\nu_ee}^{SM}=9.3\times E_{\nu_e}({\rm MeV})\times10^{-45} {\rm cm}^2.
\end{equation} 

The $\nu_e+e^-\rightarrow\nu_e+e^-$ cross section can be separated into 
its component parts: $CC$, $NC$ and interference.
Solving for the interference term and substituting the SM calculated values 
for NC and CC cross sections, the interference term can be written as 
\begin{eqnarray}
I&=&\sigma^I/2\sigma_0  \nonumber \\
 &=&\frac{\sigma_{exp}-\sigma^{CC}-\sigma^{NC}}{2\sigma_0} 
  =\frac{\sigma_{exp}-4\sigma_0-0.37\sigma_0}{2\sigma_0} \nonumber \\
 &=&\frac{\sigma_{exp}}{2\sigma_0}-2.18, 
\end{eqnarray}
where $\sigma_0=\frac{2m_eG_F^2}{4\pi}E_{\nu_e}=(4.31\times10^{-45})$ 
cm$^2$/MeV$\times E_{\nu_e}$ and becomes
\begin{equation}
I^{LSND}=-1.01\pm0.13(stat.)\pm0.12(syst.).
\end{equation}
This compares well with $I^{E225}=-1.07\pm0.17(stat.)\pm0.11(syst.)$ 
and $I^{SM}=-1.09$.

In the SM
\begin{equation}
\sigma_{\nu_ee}^{SM}=\sigma_0[1+4\sin^2\theta_W+\frac{16}{3}\sin^4\theta_W].
\end{equation}
Setting $\sigma_{\nu_ee}^{SM}=\sigma_{exp}$ we obtain
\begin{equation}
\sin^2\theta_W=0.248\pm0.051,
\end{equation}
where the error combines the statistical and systematic uncertainties.
This is in good agreement with other, much more precise measurements 
of $\sin^2\theta_W$.

Limits on the electron-neutrino charge radius were obtained following 
closely the notation and procedure of Refs. \cite{RCA93} and \cite{Allen91}.
The measured value of $\sin^2\theta_W=0.248\pm0.051$ agrees well with 
the value predicted from high energy collider results, $\sin^2\bar{\theta}_W$.
This agreement is used to place limits on the size of the radiative correction 
($\delta=\sin^2\theta_W-\sin^2\bar{\theta}_W$) 
to the electron vector coupling constant $g_V=\bar{g}_V+2\delta$ 
with $g_V=-\frac{1}{2}+2\sin^2\theta_W$ 
and $\bar{g}_V=-\frac{1}{2}+2\sin^2\bar{\theta}_W$.
The $90\%$ confidence level interval for $g_V$ 
(based on $0.159<\sin^2\theta_W<0.329$) is measured to be $-0.182<g_V<0.158$.
The radiative correction $\delta$ to the vector coupling ($\bar{g}_V=-0.04$) 
is therefore in the range $-0.142<2\delta<0.198$ 
at the $90\%$ confidence level.
Then using the relation $\delta=(\sqrt{2}\pi\alpha/3G_F)\langle r^2\rangle$ 
we obtain limits on the electron-neutrino charge radius:
\begin{equation}
-2.97\times10^{-32}<\langle r^2\rangle<4.14\times10^{-32} {\rm cm}^2
\end{equation}
A more general interpretation of these results is that they place limits 
on certain nonstandard contributions to neutrino scattering\cite{PDG98}.

It is also possible to search for neutrino flavor-changing neutral currents.
If the neutrino emerging from the neutral current differs 
from the electron-flavor neutrino emerging from the charge current, 
the two amplitudes will not add coherently.
Following Okun\cite{Okun85}, we introduce diagonal ($f_{ee}$) and 
off-diagonal ($f_{e\mu}$ and $f_{e\tau}$) couplings for neutral-current mixing 
of neutrino flavors with $1=f_{ee}^2+f_{e\mu}^2+f_{e\tau}^2$.
Non-zero flavor-changing couplings $f_{e\mu}$ or $f_{e\tau}$ would cause 
the diagonal coupling to be less than unity.
Limits on $1-f_{ee}$ can be obtained by comparing the measured cross section, 
$\sigma_{exp}$, for $\nu_e$ electron elastic scattering 
with the standard model cross section, $\sigma^{SM}$.
From the relation 
\begin{equation}
1-f_{ee}=\frac{(\sigma_{exp}-\sigma^{SM})}{4\sigma_0(1-2\sin^2\theta_W)},
\end{equation}
we obtain $1-f_{ee}<0.32$ at $90\%$ confidence level for the allowed region, 
$f_{ee}\leq1$.
Alternatively, $f_{e\mu}^2+f_{e\tau}^2<0.54$ at $90\%$ confidence level.
The E225 experiment at Los Alamos obtained similar limits\cite{RCA93,Allen92}.

Limits on the $\nu_e$ and $\nu_\mu$ magnetic moments are obtained by 
comparing the observed number of elastic events 
from $\nu_e,~\nu_\mu$ and $\bar{\nu}_\mu$ scattering, 242, 
with the 229 events expected from the Standard Model.
At $90\%$ confidence level there are then fewer than 55 events 
due to magnetic scattering.
Using the $\nu_e,~\nu_\mu$ and $\bar{\nu_\mu}$ fluxes, 
the experimental detection efficiencies and equation (\ref{eq:xsec_em}), 
the cross section for electromagnetic scattering, we obtain 
\begin{equation}
\mu_{\nu_e}^2+2.4\mu_{\nu_\mu}^2<1.1\times10^{-18}\mu_{Bohr}^2.
\end{equation}
Thus $\mu_{\nu_e}<1.1\times10^{-9}\mu_{Bohr}$ 
and $\mu_{\nu_\mu}<6.8\times10^{-10}\mu_{Bohr}$.
The limit on the muon-neutrino magnetic moment is slightly more stringent 
than that given by previous experiments\cite{LAA90,RCA93}.

\section{Conclusions}
\label{sec:conclusion}

We have measured $\nu_ee^-\rightarrow\nu_ee^-$ elastic scattering 
with a sample of $191\pm22$ events.
The reaction is of interest primarily because it is one of the few reactions 
for which the SM predicts a large destructive interference 
between the CC and NC channels.
The measured cross section, 
$\sigma_{\nu_ee^-}=[10.1\pm1.1(stat.)\pm1.0(syst.)]\times 
E_{\nu_e}($MeV$)\times10^{-45}$ cm$^2$, 
is in good agreement with standard model expectations.
The measured interference term, 
$I^{LSND}=-1.01\pm0.13(stat.)\pm0.12(syst.)$, is 
in good agreement with the SM expectation of $I^{SM}=-1.09$.
Limits are placed on neutrino flavor-changing neutral currents and 
on the electron-neutrino charge radius.
Finally, we obtain limits on the muon-neutrino magnetic moment 
that are slightly more stringent than those of previous experiments.

~~~~~

{\it Acknowledgments} 
This work was conducted under the auspices of the US Department of Energy, 
supported in part by funds provided by the University of California 
for the conduct of discretionary research by Los Alamos National Laboratory.
This work also supported by the National Science Foundation.
We particularly grateful for the extra effort that was made 
by these organizations to provide funds for running the accelerator 
at the end of the data taking period in 1995.
It is pleasing that a number of undergraduate students from participating 
institutions were able to contribute significantly to the experiment.

\begin{table}
\centering
\caption{The electron selection criteria and corresponding efficiencies for
the reaction $\nu_e+e\rightarrow\nu_e+e$.}
\begin{tabular}{ccc}
\hline
   Quantity            &    Criteria           &   Efficiency    \\    
\hline
Fiducial volume        & $D>35$ cm,            & 0.918$\pm$0.055 \\
                       & $y>-120$ cm           &                 \\
Electron energy        & $18<E_e<50$ MeV       & 0.442$\pm$0.010 \\
Particle ID            & $\chi_\alpha<4$       & 0.940$\pm$0.018 \\
In-time veto           & $<4$ PMTs             & 0.988$\pm$0.010 \\
Past activity          & See text              & 0.635$\pm$0.012 \\
Future activity        & $\Delta t_f>8.8~\mu$s & 0.991$\pm$0.003 \\
Future beta            & See Text              & 0.997$\pm$0.003 \\
Lookback               & likelihood            & 0.994$\pm$0.004 \\
DAQ and tape dead time &   --                  & 0.958$\pm$0.010 \\
Direction              & $\cos\theta>0.9$      & 0.832$\pm$0.020 \\
\hline
Total                  &                       & 0.187$\pm$0.014 \\
\hline
\end{tabular}
\label{ta:elecsel}
\end{table}

\begin{table}
\centering
\caption{Criteria to select $e^+$ from $N_{g.s.}$ beta decay and corresponding 
efficiencies for the reaction $^{12}C(\nu_e,e^-)^{12}N_{g.s.}$.}
\begin{tabular}{ccc}
\hline
Quantity            & Criteria            &    Efficiency   \\
\hline
$\beta$ decay time  & 52 $\mu$s$<t<60$ ms & 0.974$\pm$0.002 \\ 
Spatial correlation & $\Delta r<0.7$ m    & 0.992$\pm$0.008 \\
PMT threshold       & $>100$ for 1994,    & 0.875$\pm$0.011 \\
                    & $>75$ after 1994    &                 \\
Fiducial volume     & $D>0$ cm            & 0.986$\pm$0.010 \\
Trigger veto        & $>15.1~\mu$s        & 0.760$\pm$0.010 \\
Intime veto         & $<4$ PMTs           & 0.988$\pm$0.001 \\
DAQ dead time       &                     & 0.977$\pm$0.010 \\
\hline
Total               &                     & 0.612$\pm$0.015 \\ 
\hline
\end{tabular}
\label{ta:betaeff}
\end{table}

\begin{table}
\centering
\caption{Events, backgrounds and efficiency for $\cos\theta>0.9$. 
The neutrino flux and the flux averaged cross section for the reaction 
$\nu_e+e^-\rightarrow\nu_e+e^-$ are also shown.}
\begin{tabular}{cc}
\hline
Beam-on events                    & 434 events \\
Beam-off events$\times$duty ratio & 133 events \\
Beam-excess events                & 301 events \\
$\nu C$ background                & 59 events  \\
$\nu_\mu e$ background            & 24 events  \\
$\bar{\nu}_\mu e$ background      & 27 events  \\
\hline
$\nu_ee$ elastic                  & 191 events \\
Efficiency                        & 0.187      \\
$\nu_e$ flux                      & $11.76\times10^{13}$ /cm$^2$ \\
$\langle\sigma\rangle$            & $(3.19\pm0.35\pm0.33)\times10^{-43}$ cm$^2$ \\
\hline
\end{tabular}
\label{ta:events}
\end{table}

\end{document}